\documentclass[cpp,a4paper,fleqn%
]{w-art}
\usepackage{times,cite,w-thm,amsmath,amssymb}
\usepackage{epsf}
\usepackage{bm}
\theoremstyle{plain}

\theoremstyle{definition}

\usepackage[]{graphicx}

\newcommand{\Zmid}{\left<Z\right>}
\newcommand{\ZZmid}{\left<Z^2\right>}

\begin{document}
\DOIsuffix{theDOIsuffix}
\Volume{XX} \Month{XX} \Year{XXXX} 
\pagespan{1}{}
\Receiveddate{XXXX}
\Reviseddate{XXXX}
\Accepteddate{XXXX}
\Dateposted{XXXX}



\title[Plasma screening and corrections to linear mixing]{Corrections to linear mixing in binary ionic mixtures
and plasma screening at zero separation}


\author[A.\ I.\ Chugunov]{Andrey I.\ Chugunov \inst{1,}%
  \footnote{Corresponding author\quad E-mail:~\textsf{andr.astro@mail.ioffe.ru},
            Phone: +7\,812\,292\,7180,
            Fax: +7\,812\,297\,1017}}
\address[\inst{1}]{Ioffe Institute, Politekhnicheskaya 26,
194021 St.Petersburg, Russia } 
\author[H.E. DeWitt]{Hugh E. DeWitt\inst{2}}
\address[\inst{2}]{Lawrence Livermore National Laboratory, Livermore,
California 94550, USA} 
\begin{abstract}
Using the results of extensive Monte Carlo simulations we discuss
corrections to the linear mixing rule in strongly coupled binary
ionic mixtures.  We analyze the plasma screening function at zero
separation, $H_\mathit{jk}(0)$, for two ions (of types $j=1,2$ and
$k$=1,2) in a strongly coupled binary mixture. The function
$H_\mathit{jk}(0)$ is estimated by two methods: (1) from the
difference of Helmholtz Coulomb free energies at large and zero
separations; (2) by fitting the Widom expansion of
$H_\mathit{jk}(x)$ in powers of interionic distance $x$ to Monte
Carlo data on the radial pair distribution function $g_{jk}(x)$.
These methods are shown to be in good agreement. For illustration,
we analyze the plasma screening enhancement of nuclear burning rates
in dense stellar matter.
\end{abstract}
\maketitle                   

\section{Introduction}

More than 30 years ago \cite{HV76} the linear mixing rule for
multicomponent strongly coupled mixtures was shown to be highly
accurate. However, only recent studies \cite{PCR09,Mixt_New} have
achieved enough accuracy to describe the corrections to the linear
mixing rule for a wide range of plasma parameters; previous
attempts, e.g.\ \cite{DWSC96,DWS03}, were restricted at least by a
limited number of data points. We discuss the corrections to the
linear mixing rule in application to the plasma screening of nuclear
reactions in strongly coupled mixtures. Following Ref.\ \cite{DWS99}
we apply two approaches to calculate the screening enhancement: one
is based on the thermodynamic relations and the other on fitting the
mean-field potentials.
The main
advance of the present work is in using a much wider set of
numerical data and most
precise thermodynamic results.

%
\section{Plasma screening enhancement of nuclear reaction rates}\label{Sec:TwoApproaches}

Let us study a multicomponent mixture of ions $j=1, 2, \ldots$ with
atomic mass numbers $A_j$ and charge numbers $Z_j$. The ions are
supposed to be fully ionized.
Their total number density is the sum of partial densities,
$n_\mathrm{i}=\sum_j\, n_j$. It is useful to introduce the
fractional number $x_j=n_j/n_\mathrm{i}$ of ions $j$. Let us also
define the average charge number
 $\langle Z \rangle=\sum_j\, x_j Z_j$
and mass number
 $\langle A \rangle=\sum_j\, x_j A_j$
of the ions. The charge neutrality implies that the electron number
density is $n_\mathrm{e}=\langle Z \rangle n_\mathrm{i}$.
The electron plasma
screening is typically weak and will be neglected.

Thermonuclear reactions in stars take place after the atomic nuclei
collide and penetrate through the Coulomb barrier. For a not too
cold and dense stellar matter the tunneling length $r_\mathrm{t}$ is
much smaller than interionic distances (for recent result of nuclear
fusion with large tunneling distances see
\cite{OCP_react,BIM_react}). The interaction of the reacting ions
with neighboring plasma particles creates a potential well which
enlarges the number of close encounters and enhances the reaction
rate. Before the tunneling event the reactants $j$ and $k$ behave as
classical particles. Their correlations can be described by the
classical radial pair distribution function $g_{jk}(r)$. It can be
calculated by the classical Monte Carlo technique and written as
%
$    g_{jk}(r)=\exp\left[-\Gamma_{jk}\,a_{jk}/r+H_{jk}(r)/T\right]$,
%
where $\Gamma_{jk}=Z_j Z_ke^2/(a_{jk}T)$ is the correponding Coulomb
coupling parameter, and $T$ is the temperature.
The ion sphere radius $a_{jk}$ can be defined as \cite{ikm90}
$a_{jk}=(a_j+a_k)/2$ and $a_j=Z_j^{1/3} a_\mathrm{e}$, where
$a_\mathrm{e}=(4\pi n_\mathrm{e}/3)^{-1/3}$. The function
$H_{jk}(r)$ is the mean-field plasma potential. The plasma
enhancement factor is then given by
%
$
F_{jk}(r_\mathrm{t})=g_{jk}(r_\mathrm{t})/g^\mathrm{id}_{jk}(r_\mathrm{t})
    =\exp\left[H_{jk}(r_\mathrm{t})/T\right]\approx \exp\left[H_{jk}(0)/T\right].
$
%
Here,
$g^\mathrm{id}_{jk}(r_\mathrm{t})=\exp\left(-\Gamma_{jk}\,a_{jk}/r_\mathrm{t}\right)$
is the pair distribution function in the absence of screening. In
the last equality we neglect variations of $H_{jk}(r)$ over scales
$\sim r_\mathrm{t}$ which are much lower than scales $\sim a_{jk}$
of $H_{jk}(r)$.

\paragraph{Widom expansion.}

The enhancement factor of nuclear reaction rates can be determined
in following way: one can calculate $g_{jk}(r)$ by classical
Monte Carlo, extract $H_{jk}(r)$ 
and extrapolate the results to $H_{jk}(0)$. The extrapolation is
delicate \cite{rosenfeld96} because of poor Monte Carlo statistics
at small separations. We expect that the expansion of $H_{jk}(r)$
contains only even powers of $r/a_{jk}$ (the Widom expansion,
\cite{widom63}); its quadratic term is known \cite{oii91}:
\begin{equation}
   H_{jk}(r)=H_0-\frac{Z_j Z_k e^2
    }{2a^\mathrm{comp}_{jk}}
      \left(\frac{r}{a^\mathrm{comp}_{jk}}\right)^2
      +H_4\,\left(\frac{r}{a_{jk}}\right)^4
      -H_6\,\left(\frac{r}{a_{jk}}\right)^6
      +\ldots
      \label{widom}
\end{equation}
Here, $H_0=H_{jk}(0)$ and $a^\mathrm{comp}_{jk}=(Z_j+Z_k)^{1/3}
a_\mathrm{e}$ is the ion-sphere radius of the compound nuclei. Let
us also introduce the dimensionless parameter
$h^0_{jk}=H_{jk}(0)/T$. We have performed a large number of Monte
Carlo simulations of mean field potentials in binary ionic mixtures.
For each simulation, we fit $H_{jk}(r)$ by Eq.\ (\ref{widom}) taking
$H_0$, $H_4$ and $H_6$ as free parameters.
To estimate error bars we have varied $H_0$ and made additional fits
with two free parameters, $H_4$ and $H_6$.
Fig.\ \ref{Fig_Compare} shows the normalized enhancement parameter
$h_{jk}^0/\Gamma_{jk}$ (dots with error bars)
calculated in this way.

\paragraph{Thermodynamic enhancement factors.} 

The second approach to calculate the enhancement factors comes from
thermodynamics.
One can estimate
$H_{jk}(0)$ as a difference of the Helmgoltz Coulomb free energies
$F$ of the system before and after the reaction event (e.g.,
\cite{ys89}):
%
\begin{equation}
  h_{jk}^0=\left[F(\ldots,N_j,N_k,N_{jk}^\mathrm{comp},\ldots)
         -F(\ldots,N_j-1,N_k-1,N_{jk}^\mathrm{comp}+1,\ldots)\right]/T,
         \label{h0}
\end{equation}
where $N_j$, $N_k$, $N_{jk}^\mathrm{comp}$ are the numbers of the
reacting nuclei and the compound nuclei $(Z_j+Z_k,A_j+A_k)$.

Usually (see, e.g.\ \cite{BIM_react}) one assumes the linear mixing model
and presents the free energy of the Coulomb mixture $F$ as
%
$    F^\mathrm{lin}\left(\left\{N_j\right\}\right)=T\sum_j N_j
    f_0\left(\Gamma_{jj}\right),
$
%
where $f_0(\Gamma)$ is the Coulomb free energy (normalized to temperature
$T$) per one nucleus in one component plasma.
We use the well known approximation of
$f_0(\Gamma)$
suggested by Potekhin \& Chabrier
\cite{pc00}.
In linear mixing model Eq.\ (\ref{h0}) can be written in
the convenient form:
\begin{equation}
 h_{jk}^\mathrm{lin}=f_0(\Gamma_{jj})+f_0(\Gamma_{kk})-f_0(\Gamma_{jk}^\mathrm{comp}),
 \label{h0lin}
\end{equation}
where
$\Gamma_{jk}^\mathrm{comp}=\left(Z_j+Z_k\right)^{5/3}\Gamma_\mathrm{e}$
is the Coulomb coupling parameter for the compound nucleus. The values
of   $h_{jk}^\mathrm{lin}$ are shown by the solid line in
Fig.\ \ref{Fig_Compare}.

Our aim is to check the accuracy of the linear mixing and analyze
deviations from this model.
To do this we apply the best available results for the
thermodynamics of multicomponent mixtures \cite{PCR09,Mixt_New},
which are valid for any value of the coupling parameter. The values
of the corresponding enhancement parameter $h_{jk}^0/\Gamma_{jk}$
are shown by the long-dash line in Fig.\ \ref{Fig_Compare}.

\section{Comparison of different approaches}\label{Sec:compare}
\begin{figure}
    \begin{center}
        \leavevmode
        \epsfxsize=155mm \epsfbox{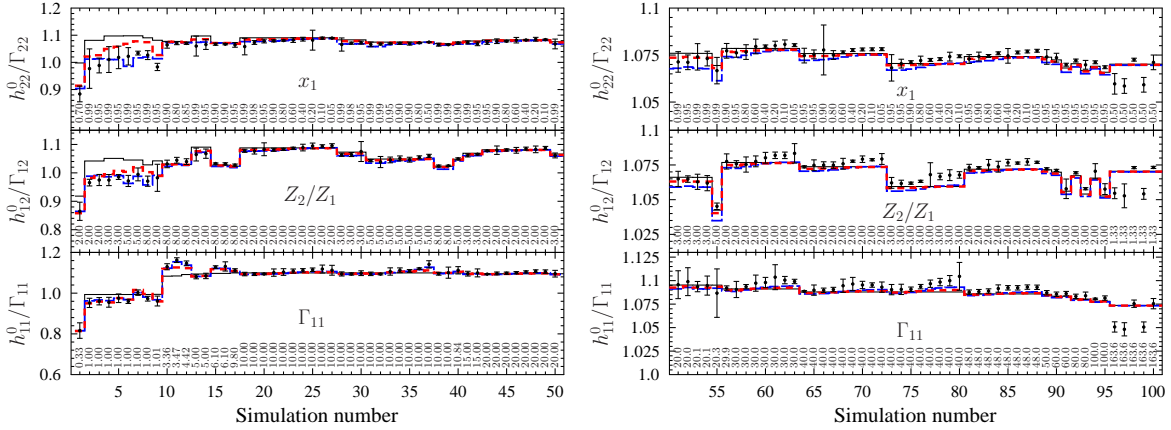}
    \end{center}
    \caption{(Color online)
    Histogram of the enhancement factors
    extracted from:
    1) Widom fitting 
    (dots with error bars);
    2) linear mixing [Eq.\ (\ref{h0lin}); solid line];
    3) thermodynamics 
    (long-dash lines);
    4) our approximation [Eq.\ (\ref{appr}); short-dash lines].
      }
    \label{Fig_Compare}
\end{figure}

In Fig.\  \ref{Fig_Compare} we compare the plasma screening function at zero
separation calculated by different methods.
Each of six panels demonstrates the histogram of normalized
screening functions $h_{jk}^0/\Gamma_{jk}$ versus simulation number.
Three left panels show simulations with numbers from 1 to 50, and
three right panels show simulations from 51 to 100. For each 3-panel
block, the lower panel presents $h^0_{11}/\Gamma_{11}$, the middle
panel shows $h_{12}^0/\Gamma_{12}$, and the upper panel gives
$h^0_{22}/\Gamma_{22}$. The parameters of simulations
$(\Gamma_{11},\ Z_2/Z_1,\ x_1)$ are also shown on each block by
vertically aligned numbers: $\Gamma_{11}$ on the lower panel,
$Z_2/Z_1$ on the middle and $x_1$ on the upper panel. For example,
the simulation number 1 
is done
for $\Gamma_{11}\approx0.33$, $Z_2/Z_1=2$, and $x_1=0.7$.

Each panel contains a set of dots with error bars, which represent
the values of $h_{jk}^0/\Gamma_{jk}$ calculated by fitting
$H_{jk}(r)$ with the aid of (\ref{widom}). Each panel contains 3
lines: the solid line shows the results of the linear mixing model,
Eq.\ (\ref{h0lin}); the long-dash line is calculated with the best
available thermodynamics of the multicomponent plasma (Eq.\
(\ref{h0}) with the free energy taken from \cite{Mixt_New}); the
short-dash line is our approximation (\ref{appr}). Note, that the
normalized enhancement parameter is approximately constant at large
$\Gamma_{jk}$. This property is well known \cite{salpeter54}.
The linear mixing is a highly accurate as long as
$\Gamma_{jk}\gtrsim 10$. For lower $\Gamma_{jk}$ the relative
corrections can be much larger and well described by both
dashed-lines (the accurate thermodynamics and approximation). The
most noticeable difference between dots and the short-dashed lines
takes place for $h_{22}/\Gamma_{22}$ in simulations 6, 7, 8, and 9
that are done for low fractions of highly charged ions and large
ratio $Z_2/Z_1\ge5$. Such a difference is unimportant for
applications
--- it translates into the correction to the reaction rate within
a factor of two.

Also, there are three large $\Gamma$ simulations (96, 97, and 99),
where dots are divergent. They started with lattice configurations
of ions. Thus
the corrections to the linear mixing in crystalline phase are larger
(as noted in \cite{DWS03}).

\section{Approximation of enhancement factors and conclusions}\label{Sec:approx}

We suggest to use the following approximation for the enhancement
factor for all $\Gamma$ and mixture composition
\begin{equation}
   h^0_{jk}=h^\mathrm{lin}_{jk}/
          \left[1
               +C_{jk}\left(1-C_{jk}\right)\,
                \left(h^\mathrm{lin}_{jk}/h^\mathrm{DH}_{jk}
                \right)^2
          \right].
          \label{appr}
\end{equation}
Here, $h^\mathrm{lin}_{jk}$ is given by (\ref{h0lin}),
%
 $ h^\mathrm{DH}_{jk}=3^{1/2}
  Z_j\,Z_k\ZZmid^{1/2}\Gamma_\mathrm{e}^{3/2}/\Zmid^{1/2}$ is
the well known Debye-H\"{u}ckel enhancement parameter, and   
  $C_{jk}=
   3Z_j\,Z_k\ZZmid^{1/2}\Zmid^{-1/2}/\left[\left(Z_j+Z_k\right)^{5/2}-Z_j^{5/2}-Z_k^{5/2}\right]
  $.
Eq.\ (\ref{appr}) reproduces the Debye-H\"{u}ckel asymptote at low
$\Gamma$ and the linear mixing at strong coupling.

\begin{figure}
    \begin{center}
        \leavevmode
        \epsfxsize=150mm \epsfbox{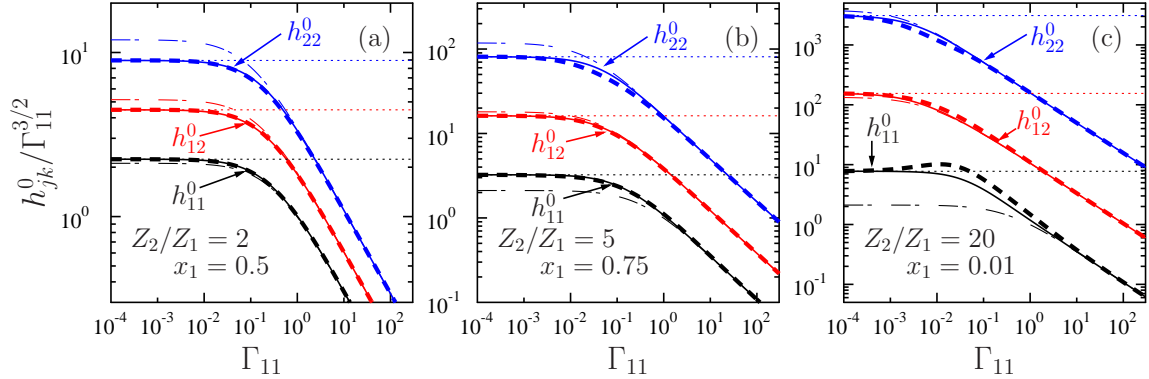}
    \end{center}
    \caption{(Color online) Enhancement factors $h_{jk}^0/\Gamma_{11}^{3/2}$ vs $\Gamma_{11}$
    for three binary ionic mixtures.
      }
    \label{Fig_hvsG}
\end{figure}

In Fig.\ \ref{Fig_hvsG} we show the dependence of the approximated
enhancement factors $h_{jk}^0/\Gamma_{11}^{3/2}$ on $\Gamma_{11}$.
The figure contains three panels; each for a specific binary ionic
mixture. Each panel shows three groups of four lines. They are (from
top to bottom) $h^0_{22}/\Gamma_1^{3/2}$, $h^0_{12}/\Gamma_1^{3/2}$
and $h_{11}^0/\Gamma_1^{3/2}$. Two of any four lines (solid and
thick dashed lines) are almost the same in the majority of cases.
This couple represents the approximation (\ref{appr}) and the
thermodynamic enhancement factor (\ref{h0}), respectively. The
dotted horizontal lines refer to the Debye-H\"{u}ckel model and the
dash-dot lines are the linear mixing results. One can see that our
approximation is in a good agreement with thermodynamic results for
most of cases, especially in panel (a) (for all mixtures with not
too large $Z_2/Z_1$). If $Z_2/Z_1$ becomes too large [panel (c)],
the thermodynamic model of $h_{11}^0/\Gamma_1$, calculated in
accordance with \cite{Mixt_New}, has a specific feature
($h_{11}/\Gamma_{11}^{3/2}$ increases at $\Gamma_{11}\sim10^{-2}$),
while our approximation has not. We expect that this feature is not
real, but results from not too accurate extractions of the
enhancement factors from thermodynamic data. The free energy is
almost fully determined by larger charges $Z_2$ which also dominate
by number ($99\%$) in panel (c). Using Eq.\ \ref{h0} to get
$h_{11}^0$, one should differentiate the free energy with respect to
$N_1$, which provides vanishing contribution to the free energy.
Hence this procedure is very delicate and can strongly amplify the
errors of original thermodynamic approximation. We expect that our
approximation can be more accurate than the original thermodynamic
result. Another, less probable option is that we still have not
enough data to prove the presence of the feature of
$h^0_{11}/\Gamma_{11}^{3/2}$.

%
%
%
To conclude, we have calculated the enhancement factors of nuclear
reactions in binary ionic mixtures by two methods and showed good
agreement of the results. We have proposed a simple approximation of
the enhancement factors valid for any Coulomb coupling. This
approximation is almost the same as thermodynamic ones for not too
specific mixtures. It does not confirm some questionable features of
the enhancement factors for mixtures with large $Z_2/Z_1$ and small
$x_1$.

\begin{acknowledgement}
 We are grateful to D.G.~Yakovlev and A.Y.~Potekhin for useful remarks. Work of AIC was
partly supported by the Russian Foundation for Basic Research (grant
08-02-00837), and by the State Program ``Leading Scientific Schools
of Russian Federation'' (grant NSh 2600.2008.2). Work of HED was
performed under the auspices of the US Department of Energy by the
Lawrence Livermore National Laboratory under contract number
W-7405-ENG-48.
\end{acknowledgement}

\end{document}